\documentclass[12pt]{article}
\usepackage[latin1]{inputenc}
\usepackage[british]{babel}
\usepackage{cmap}
\usepackage{lmodern}

\usepackage{amssymb, amsmath, amsthm}
\usepackage[a4paper,top=25mm,bottom=25mm,left=25mm,right=25mm]{geometry}
\usepackage{ragged2e}

\usepackage{authblk} 
\usepackage{pifont}
\usepackage{graphicx}
\usepackage[usenames,dvipsnames,svgnames,table]{xcolor}
\usepackage[figuresright]{rotating}
\usepackage{xtab} 
\usepackage{longtable} 
\usepackage{multirow}
\usepackage{footnote}
\usepackage[stable]{footmisc}
\usepackage{chngpage} 
\usepackage{pdflscape} 
\usepackage[nottoc,notlot,notlof]{tocbibind} 

\usepackage{pgfplots}
\pgfplotsset{compat=1.14}
\pgfplotsset{every tick label/.append style={font=\footnotesize}}
\usepackage{setspace}

\makesavenoteenv{tabular}
\usepackage{tabularx}
\usepackage{booktabs}
\usepackage{threeparttable} 
\usepackage[referable]{threeparttablex} 
\newcolumntype{R}{>{\raggedleft\arraybackslash}X}
\newcolumntype{L}{>{\raggedright\arraybackslash}X}
\newcolumntype{C}{>{\centering\arraybackslash}X}
\newcolumntype{A}{>{\columncolor{gray!25}}C}
\newcolumntype{a}{>{\columncolor{gray!25}}c}

\newlength{\tablen}

\usepackage{dcolumn} 
\newcolumntype{.}{D{.}{.}{-1}}

\usepackage{tikz}
\usetikzlibrary{arrows, calc, matrix, patterns, positioning, shapes, trees}
\usepackage[semicolon]{natbib}
\usepackage[hyphens]{url}
\usepackage{hyperref} 
\hypersetup{
  colorlinks   = true,    
  urlcolor     = blue,    
  linkcolor    = blue,    
  citecolor    = ForestGreen      
}
\usepackage{microtype}
\usepackage[singlelinecheck=false,justification=centering]{caption} 

\usepackage[labelformat=simple]{subcaption}

\DeclareCaptionLabelFormat{parenthesis}{(#2)}
\makeatletter
\renewcommand\p@subfigure{\arabic{figure}.}
\makeatother

\DeclareCaptionLabelFormat{parenthesis}{(#2)}
\makeatletter
\renewcommand\p@subtable{\arabic{table}.}
\makeatother

\usepackage{enumitem}
\setlist[itemize]{leftmargin=2.5\parindent}
\setlist[enumerate]{leftmargin=2.5\parindent}

\def\addlegendimage{\csname pgfplots@addlegendimage\endcsname}

\theoremstyle{plain}

\theoremstyle{definition}

\theoremstyle{remark}

\def\keywords{\vspace{.5em} 
{\noindent \textit{Keywords}: }}

\def\JEL{\vspace{.5em} 
{\noindent \textbf{\emph{JEL} classification number}: }}

\def\AMS{\vspace{.5em} 
{\noindent \textbf{\emph{MSC} class}: }}

\title{Group draw with unknown qualified teams: \\ A lesson from the 2022 FIFA World Cup draw}
\author{\href{https://sites.google.com/view/laszlocsato}{L\'aszl\'o Csat\'o}\thanks{~E-mail: \emph{laszlo.csato@sztaki.hu}} }
\affil{Institute for Computer Science and Control (SZTAKI) \\
E\"otv\"os Lor\'and Research Network (ELKH) \\
Laboratory on Engineering and Management Intelligence \\
Research Group of Operations Research and Decision Systems}
\affil{Corvinus University of Budapest (BCE) \\
Department of Operations Research and Actuarial Sciences}
\affil{Budapest, Hungary}
\date{\today}

\def\Dedication{
{\noindent
$\mathfrak{Wir}$ $\mathfrak{behaupten}$ $\mathfrak{aber}$, $\mathfrak{da\ss}$, $\mathfrak{wo}$ $\mathfrak{es}$ $\mathfrak{auf}$ $\mathfrak{die}$ $\mathfrak{Feststellung}$ $\mathfrak{einer}$ $\mathfrak{neuen}$ $\mathfrak{oder}$ $\mathfrak{einer}$ $\mathfrak{zweifelhaften}$ $\mathfrak{Meinung}$ $\mathfrak{ankommt}$, $\mathfrak{ein}$ $\mathfrak{einziges}$ $\mathfrak{gr\ddot{u}ndlich}$ $\mathfrak{dargestelltes}$ $\mathfrak{Ereignis}$ $\mathfrak{belehrender}$ $\mathfrak{ist}$ $\mathfrak{als}$ $\mathfrak{zehn}$ $\mathfrak{blo\ss}$ $\mathfrak{ber\ddot{u}hrte}$.\footnote{~``\emph{But we maintain that when the object is to establish a new or doubtful opinion, one single example, thoroughly analysed, is far more instructive than ten which are superficially treated.}'' (Source: Carl von Clausewitz: \emph{On War}, Book 2, Chapter 6 [On Examples]. Translated by Colonel James John Graham, London, N. Tr\"ubner, 1873. \url{http://clausewitz.com/readings/OnWar1873/TOC.htm})}
}
\vspace{0.25cm}

\flushright
\noindent (Carl von Clausewitz: \emph{Vom Kriege})

\vspace{1cm} 
\justify }

\begin{document}

\newgeometry{top=25mm,bottom=25mm,left=25mm,right=25mm}

\maketitle
\thispagestyle{empty}
\Dedication

\begin{abstract}
\noindent
The draw for the 2022 FIFA World Cup has been organised before the identity of three winners of the play-offs is revealed. Seeding has been based on the FIFA World Ranking released on 31 March 2022 but these three teams have been drawn from the weakest Pot 4. We show that the official seeding policy does not balance the difficulty levels of the groups to the extent possible: a better alternative would have been to assign the placeholders according to the highest-ranked potential winner, similar to the rule used in the UEFA Champions League qualification.
Our simulations reinforce that this is the best strategy in general to create balanced groups in the FIFA World Cup.


\AMS{90-10, 90B90, 91B14}

\JEL{C44, C63, Z20}
\end{abstract}

\clearpage
\restoregeometry

\section{Introduction} \label{Sec1}

Inspired by the criticism of \citet{Guyon2014c} and \citet{Guyon2015a}, FIFA has reformed the draw of the 2018 World Cup in order to produce balanced groups \citep{Guyon2018d, CeaDuranGuajardoSureSiebertZamorano2020}: according to the classical scheme, the 32 national teams have been divided into four pots based on the FIFA World Ranking (except for favouring the host Russia by assigning it to the strongest pot), and each group has consisted of a team from each group.
However, because of the COVID-19 pandemic and the 2022 Russian invasion of Ukraine, FIFA has been forced to draw the groups of the 2022 World Cup when the identity of three teams has remained unknown. The draw has taken place on 1 April 2022, while the two inter-confederation play-offs are contested in June 2022 and the qualification match(es) of Ukraine have been rescheduled to the same month \citep{FIFA2022b}.

This raises a problem since seeding is based on the FIFA World Ranking released on 31 March 2022. The Organising Committee for FIFA Competitions has decided to assign the three placeholders to Pot 4, that is, among the weakest teams.

We show that this questionable policy unnecessarily worsens the balance in the strengths of the groups. A better outcome can be provided by assigning the unknown placeholders according to the highest-ranked remaining team in each undecided contest. The proposed rule is currently used in the qualifications for the European club competitions \citep{Csato2022b}. The solution chosen by FIFA is detrimental to some national teams, including Ukraine.

To generalise our finding, three policies are compared in a stylised model of the FIFA World Cup draw with respect to the expected strengths of the groups: the placeholder from an undecided contest is assigned 
(1) according to the highest-ranked participant;
(2) according to the lowest-ranked participant; or
(3) to the weakest pot.
The first option turns out to be the best to produce balanced groups.

Naturally, our paper has antecedents in the extant literature.
The uneven distribution of the 1990 \citep{Jones1990}, 2006 \citep{RathgeberRathgeber2007}, 2014 \citep{Guyon2015a}, and 2018 FIFA World Cups \citep{Csato2022e} have already been demonstrated. \citet{RobertsRosenthal2022} suggest two practical methods that are suitable for a televised draw to guarantee uniform distribution. \citet{CeaDuranGuajardoSureSiebertZamorano2020}, \citet{Guyon2015a}, and \citet{LalienaLopez2019} have proposed draw systems for sports tournaments in the presence of geographical or seeding restrictions to create balanced groups with roughly the same competitive level. However, all these suggestions require a fundamentally new draw procedure, which is unlikely to be adopted soon. On the other hand, our recommendation builds on a principle already used by the Union of European Football Associations (UEFA).

The remainder of the work is structured as follows. Section~\ref{Sec2} summarises the rules of the 2022 FIFA World Cup draw. The methodology of our analysis is described in Section~\ref{Sec3}, and the findings are presented in Section~\ref{Sec4}. Section~\ref{Sec5} attempts to derive general results, while Section~\ref{Sec6} concludes.

\section{Draw systems for the 2022 FIFA World Cup} \label{Sec2}

This section describes the draw procedure that has been used in the 2022 FIFA World Cup draw on 1 April 2022 \citep{FIFA2022a}. It determines the allocation of 29 qualified teams, the winners of two inter-confederation play-offs, and the placeholder of a UEFA play-off slot into eight groups of four teams each. In addition, we argue for an alternative draw procedure.

\subsection{The official rules} \label{Sec21}

The 32 teams are divided into four pots on the basis of the FIFA World Ranking released on 31 March 2022 that already takes the results of qualification games played in the March 2022 international match window into account:
\begin{itemize}
\item
Pot 1 contains the host Qatar, automatically assigned to Group A, and the seven highest-ranked teams;
\item
Pot 2 contains the teams ranked 8th to 15th;
\item
Pot 3 contains the teams ranked 16th to 23rd;
\item
Pot 4 includes the teams ranked 24th to 28th plus the two placeholders from the two inter-confederation play-offs and the winner of the UEFA play-off Path A.
\end{itemize}
The inter-confederation play-offs are scheduled to be played on 13--14 June 2022 in Qatar. Two matches in the UEFA play-off Path A have been postponed out of necessity due to the Russian invasion of Ukraine as one semifinal and possibly the final involves Ukraine. 

The draw sequence starts with Pot 1 and ends with Pot 4.
Each pot is emptied before moving on to the next one. Some draw conditions apply to ensure geographic separation \citep{FIFA2022a}:
\begin{itemize}
\item
No group can have more than one team from any continental confederation except for UEFA (AFC, CAF, CONMEBOL, CONCACAF).
\item
Each group should consist of at least one but no more than two European teams.
\end{itemize}
Since the 2022 World Cup will be contested by 13 UEFA members, five out of the eight groups are guaranteed to include two teams from Europe. The allocation of the two inter-confederation play-offs is based on the confederation of both potential winners.

Even though the official overview of the draw procedure \citep{FIFA2022a} does not specify how the draw constraints are met, clearly, the standard procedure of the FIFA/UEFA \citep{Csato2022e} is used. In particular, the team drawn is placed in the first available group in alphabetical order as indicated by the computer such that any deadlock situation (when the teams still to be drawn cannot be allocated into the remaining slots without violating a draw condition) is prevented.

For instance, assume that Group F/G/H contains Senegal/Morocco/Tunisia (all CAF) from Pot 3, respectively, whereas Cameroon and Ghana (both CAF) are among the five remaining teams in Pot 4. If the next empty slot in alphabetical order is in Group D and the fourth team drawn from Pot 4 is neither Cameroon nor Ghana, the latter cannot be assigned to Group D because otherwise, two African countries should be allocated for the four available groups but three of them are prohibited for CAF teams, which is impossible.
This procedure is explained in a video available at \url{https://www.youtube.com/watch?v=jDkn83FwioA} through the example of the 2018 FIFA World Cup.
The mechanism has first been proposed in \citet{Guyon2014c} for the FIFA World Cup draw and has been adopted by FIFA in 2018 \citep{Guyon2018d}.
It has already received serious scrutiny in the literature \citep{BoczonWilson2018, Csato2022e, KlossnerBecker2013}.

\subsection{A reasonable alternative seeding policy} \label{Sec22}

The assignment of the three placeholders representing the winners of the play-offs to the weakest pot is a questionable decision since they can be relatively strong teams as we will see later.
The same problem arises in the qualification stages of the UEFA Champions League and the UEFA Europa Conference League but it is treated in a different way: ``\emph{If, for any reason, any of the participants in such rounds are not known at the time of the draw, the coefficient of the club with the higher coefficient of the two clubs involved in an undecided tie is used for the purposes of the draw}'' \citep[Article~13.03]{UEFA2021c, UEFA2021e}.
According to this policy, the placeholders of the play-offs that are still to be contested for the FIFA World Cup should be placed in a pot based on the highest-ranked potential winner instead of Pot 4.

\section{Methodology} \label{Sec3}

Due to the assignment of the host in Group A, the 2022 FIFA World Cup draw has $7 \times \left( 8! \right)^3 \approx 3.3 \times 10^{17}$ possible outcomes without accounting for geographic restrictions. Even though these criteria significantly  decrease the number of feasible solutions, it is still impossible to exactly calculate the probability of each assignment. Furthermore, the consequences of choosing a particular seeding regime can only be uncovered if the results of matches played in the play-offs and the groups are determined. To that end, computer simulations will be used as recommended in the literature on tournament design \citep{ScarfYusofBilbao2009}.

Since two teams from each group advance to the Round of 16, a group is usually judged to be tough when three teams have high rankings, even if the fourth one is much weaker \citep{Guyon2015a, LalienaLopez2019}. Hence, our measure of group strength will be the weighted average of the ratings of the four participants, where the weight of the strongest, the second strongest, and the third strongest team is two, whereas the weight of the weakest team is one.

The abilities of the teams will be quantified in two ways.
The first is the rating points in the FIFA World Ranking of March 2022, underlying the pot allocation. Although FIFA has adopted the Elo method of calculation after the 2018 FIFA World Cup \citep{FIFA2018a, FIFA2018c}, the current FIFA World Ranking does not take home advantage and the margin of victory into account. Both factors are considered in the World Football Elo Ratings (\url{http://eloratings.net}), which is a widely used benchmark in the literature \citep{CeaDuranGuajardoSureSiebertZamorano2020, Csato2022a, Guyon2014c, Guyon2015a, LasekSzlavikBhulai2013, LasekSzlavikGagolewskiBhulai2016}. This will provide the second measure for the strengths of the teams.

As the FIFA ranking is somewhat slow to react to the changing skill level of the teams (the example of Canada and Ecuador will be seen later) and is still influenced by the transformation from the old ranking method in 2018 (Ecuador has a real difficulty gaining enough points to climb substantially since it mainly plays against other South American teams), the outcomes of all matches will be simulated on the basis of the World Football Elo Ratings.
A traditional choice for the distribution of the number of goals in soccer is the Poisson distribution \citep{ChaterArrondelGayantLaslier2021, Maher1982, VanEetveldeLey2019}. Then the probability that team $i$ scores $k$ goals against team $j$ is
\begin{equation*} \label{Poisson_dist}
P_{ij}(k) = \frac{ \left( \lambda_{ij}^{(f)} \right)^k \exp \left( -\lambda_{ij}^{(f)} \right)}{k!},
\end{equation*}
where the expected number of goals scored by team $i$ against team $j$ is $\lambda_{ij}^{(f)}$ if the match is played on field $f$ (home: $f = h$; away: $f = a$; neutral: $f = n$).

\citet{FootballRankings2020} determines parameter $\lambda_{ij}^{(f)}$ as a quartic polynomial of the win expectancy $W_{ij}$ of team $i$ against team $j$, which is
\begin{equation*} \label{eq1}
W_{ij} = \frac{1}{1 + 10^{-(E_i - E_j)/400}},
\end{equation*}
with $E_i$ and $E_j$ being the Elo ratings of the two teams, respectively. The rating of the home team is increased by $100$ to reflect home advantage.
The exact formulas are estimated by a least squares regression based on more than 29 thousand home-away matches and almost 10 thousand games played on neutral ground between national football teams. In addition, they contain a regime change at $W_{ij} = 0.9$ since unbalanced games usually mean an excessive number of goals.

Most of the games are played on neutral ground when the expected number of goals for team $i$ against team $j$ is
\begin{equation*} \label{Exp_goals_neutral}
\lambda_{ij}^{(n)} = 
\left\{ \begin{array}{ll}
3.90388 \cdot W_{ij}^4 - 0.58486 \cdot W_{ij}^3 \\
- 2.98315 \cdot W_{ij}^2 + 3.13160 \cdot W_{ij} + 0.33193 & \textrm{if } W_{ij} \leq 0.9 \\ \\
308097.45501 \cdot (W_{ij}-0.9)^4 - 42803.04696 \cdot (W_{ij}-0.9)^3 & \\
+ 2116.35304 \cdot (W_{ij}-0.9)^2 - 9.61869 \cdot (W_{ij}-0.9) + 2.86899 & \textrm{if } W_{ij} > 0.9,
\end{array} \right.
\end{equation*}
with $R^2 = 0.976$.

However, there are some home-away matches (two in the UEFA play-offs and the three group matches of Qatar) to be simulated, where the expected number of goals for the home team $i$ equals
\begin{equation*} \label{Exp_goals_home}
\lambda_{ij}^{(h)} = 
\left\{ \begin{array}{ll}
-5.42301 \cdot W_{ij}^4 + 15.49728 \cdot W_{ij}^3 \\
- 12.6499 \cdot W_{ij}^2 + 5.36198 \cdot W_{ij} + 0.22862 & \textrm{if } W_{ij} \leq 0.9 \\ \\
231098.16153 \cdot (W_{ij}-0.9)^4 - 30953.10199 \cdot (W_{ij}-0.9)^3 & \\
+ 1347.51495 \cdot (W_{ij}-0.9)^2 - 1.63074 \cdot (W_{ij}-0.9) + 2.54747 & \textrm{if } W_{ij} > 0.9
\end{array} \right.
\end{equation*}
with $R^2 = 0.984$, and the expected number of goals for the away team $j$ is given by
\begin{equation*} \label{Exp_goals_away}
\lambda_{ij}^{(a)} = 
\left\{ \begin{array}{ll}
90173.57949 \cdot (W_{ij} - 0.1)^4 + 10064.38612 \cdot (W_{ij} - 0.1)^3 \\
+ 218.6628 \cdot (W_{ij} - 0.1)^2 - 11.06198 \cdot (W_{ij} - 0.1) + 2.28291 & \textrm{if } W_{ij} < 0.1 \\ \\
-1.25010 \cdot W_{ij}^4 -  1.99984 \cdot W_{ij}^3 & \\
+ 6.54946 \cdot W_{ij}^2 - 5.83979 \cdot W_{ij} + 2.80352 & \textrm{if } W_{ij} \geq 0.1
\end{array} \right.
\end{equation*}
with $R^2 = 0.955$.

The same simulation model has been used recently to quantify the incentive incompatibility of the European Qualifiers for the 2022 FIFA World Cup \citep{Csato2022a} and the unfairness of the 2018 FIFA World Cup qualification \citep{Csato2022c}.

The play-offs contain single-game matches, hence draws are not allowed. If the two teams score the same number of goals, the winner is chosen randomly. This effectively means that there is no goal in extra time and the penalty shootout provides equal chances for the two teams, independently of the field of the game.

The ranking of the teams in the groups is determined according to the following criteria:
(a) greatest number of points obtained in all group matches;
(b) goal difference in all group matches;
(c) greatest number of goals scored in all group matches;
(d) drawing of lots.

\begin{table}[t!]
  \centering
  \caption{National teams qualified for the 2022 FIFA World Cup before the draw}
  \label{Table1}
    \rowcolors{1}{}{gray!20}
\begin{threeparttable}
    \begin{tabularx}{0.8\textwidth}{LLm{2cm}m{1cm}} \toprule
    Country & Confederation & Points & Elo \\ \bottomrule
    Qatar & AFC   & 1441.41 & 1662 \\
    Brazil & CONMEBOL & 1832.69 & 2155 \\
    Belgium & UEFA  & 1827.00  & 2069 \\
    France & UEFA  & 1789.85 & 2116 \\
    Argentina & CONMEBOL & 1765.13 & 2018 \\
    England & UEFA  & 1761.71 & 2039 \\
    Spain & UEFA  & 1709.19 & 2039 \\
    Portugal & UEFA  & 1674.78 & 1984 \\ \hline
    Mexico & CONCACAF & 1658.82 & 1848 \\
    Netherlands & UEFA  & 1658.66 & 1938 \\
    Denmark & UEFA  & 1653.60 & 1936 \\
    Germany & UEFA  & 1650.53 & 1966 \\
    Uruguay & CONMEBOL & 1635.73 & 1923 \\
    Switzerland & UEFA  & 1635.32 & 1920 \\
    United States & CONCACAF & 1633.72 & 1822 \\
    Croatia & UEFA  & 1621.11 & 1855 \\ \hline
    Senegal & CAF   & 1584.16 & 1729 \\
    Iran  & AFC   & 1564.49 & 1820 \\
    Japan & AFC   & 1553.44 & 1796 \\
    Morocco & CAF   & 1551.88 & 1738 \\
    Serbia & UEFA  & 1547.43 & 1845 \\
    Poland & UEFA  & 1544.20 & 1799 \\
    South Korea & AFC   & 1519.54 & 1800 \\
    Tunisia & CAF   & 1499.80 & 1612 \\ \hline
    Cameroon & CAF   & 1480.48 & 1631 \\
    Canada & CONCACAF & 1479.00  & 1798 \\
    Ecuador & CONMEBOL & 1452.63 & 1840 \\
    Saudi Arabia & AFC   & 1444.69 & 1634 \\
    Ghana & CAF   & 1387.36 & 1541 \\ \toprule
    \end{tabularx}
\begin{tablenotes} \footnotesize
\item
Horizontal lines indicate the boundaries of the pots.
\item
The column Points shows the strength of the teams according to the FIFA World Ranking as of 31 March 2022, see \url{https://www.fifa.com/fifa-world-ranking/men?dateId=id13603}.
\item
The column Elo shows the strength of the teams according to the World Football Elo Ratings as of 31 March 2022, see \url{https://www.international-football.net/elo-ratings-table?year=2022&month=03&day=31}.
\end{tablenotes}
\end{threeparttable}
\end{table}

Table~\ref{Table1} shows the composition of the pots, as well as the two measures of strength for the teams.

\begin{table}[t!]
  \centering
  \caption{Matches for the three available FIFA World Cup slots}
  \label{Table2}
    \rowcolors{1}{}{gray!20}
\centerline{
\begin{threeparttable}
    \begin{tabularx}{1.1\textwidth}{Llcc lcc l} \toprule
    Match & Team 1 & Points & Elo   & Team 2 & Points & Elo   & Field \\ \bottomrule
    AFC PO & Australia & 1462.29 & 1677  & United Arab Emirates & 1356.99 & 1515  & Qatar \\
    ICT PO1 & \multicolumn{3}{l}{Winner of AFC PO} & Peru  & 1562.32 & 1856  & Qatar \\
    ICT PO2 & Costa Rica & 1503.09 & 1743  & New Zealand & 1206.07 & 1558  & Qatar \\
    UEFA SF & Scotland & 1472.66 & 1730  & Ukraine & 1535.08 & 1817  & Scotland \\
    UEFA PO & Wales & 1588.08 & 1841  & \multicolumn{3}{l}{Winner of UEFA SF} & Wales \\ \toprule
    \end{tabularx}
\begin{tablenotes} \footnotesize
\item
The column Points shows the strength of the teams according to the FIFA World Ranking as of 31 March 2022, see \url{https://www.fifa.com/fifa-world-ranking/men?dateId=id13603}.
\item
The column Elo shows the strength of the teams according to the World Football Elo Ratings as of 31 March 2022, see \url{https://www.international-football.net/elo-ratings-table?year=2022&month=03&day=31}.
\end{tablenotes}
\end{threeparttable}
}
\end{table}

Three play-off paths are not yet finished at the time of the draw. The corresponding matches are listed in Table~\ref{Table2}.

\begin{table}[t!]
  \centering
  \caption{The alternative seeding for the 2022 FIFA World Cup draw}
  \label{Table3}
    \rowcolors{1}{gray!20}{}
    \begin{tabularx}{\textwidth}{llCll} \toprule \hiderowcolors
    Country & Confederation &       & Country & Confederation \\ \midrule
    \multicolumn{2}{c}{\textbf{Pot 1}} &       & \multicolumn{2}{c}{\textbf{Pot 2}} \\ \bottomrule \showrowcolors
    Qatar & AFC   &       & Mexico & CONCACAF \\
    Brazil & CONMEBOL &       & Netherlands & UEFA \\
    Belgium & UEFA  &       & Denmark & UEFA \\
    France & UEFA  &       & Germany & UEFA \\
    Argentina & CONMEBOL &       & Uruguay & CONMEBOL \\
    England & UEFA  &       & Switzerland & UEFA \\
    Spain & UEFA  &       & United States & CONCACAF \\
    Portugal & UEFA  &       & Croatia & UEFA \\ \bottomrule
    \multicolumn{2}{c}{\textbf{Pot 3}} &       & \multicolumn{2}{c}{\textbf{Pot 4}} \\ \toprule
    Senegal & CAF   &       & Cameroon & CAF \\
    Iran  & AFC   &       & Canada & CONCACAF \\
    Japan & AFC   &       & Ecuador & CONMEBOL \\
    Morocco & CAF   &       & Saudi Arabia & AFC \\
    Serbia & UEFA  &       & Ghana & CAF \\
    Poland & UEFA  &       & South Korea & AFC \\
    $\mathcal{W}$/ICT PO1 & AFC/CONMEBOL &       & $\mathcal{W}$/ICT PO2 & CONCACAF/OFC \\
    $\mathcal{W}$/UEFA PO & UEFA  &       & Tunisia & CAF \\ \toprule
    \end{tabularx}
\end{table}

Finally, the alternative pot allocation is presented in Table~\ref{Table3}.

A simulation run consists of the following steps:
\begin{enumerate}
\item
The winners of the remaining matches in the play-offs are determined;
\item
The groups of the 2022 FIFA World Cup are drawn according to both seeding rules (official and alternative): the teams in each pot are ordered randomly and assigned sequentially to the first available group in alphabetical order such that all draw conditions are satisfied;
\item
The expected strength of each group is computed according to both measures (rating points in the FIFA World Ranking and World Football Elo Ratings, see Tables~\ref{Table1} and \ref{Table2});
\item
All group matches are played, group rankings and the set of qualified teams are obtained.
\end{enumerate}
All simulations are carried out 1 million times to smooth the effect of random fluctuations.

\section{Results for the 2022 FIFA World Cup draw} \label{Sec4}

In the following, our findings about the 2022 FIFA World Cup draw will be presented. In particular, Section~\ref{Sec41} addresses the balance across groups by quantifying their competitive level. The consequences of the official seeding regime with respect to the probability of qualification are uncovered in Section~\ref{Sec42}.

\subsection{Group balance} \label{Sec41}

Two seeding rules have been outlined in Section~\ref{Sec2} and two measures of group strength have been suggested in Section~\ref{Sec3}. In each simulation run, the groups have been ordered according to their strength, and the averages of these values over the 1 million iterations have been computed. Group A is treated separately since Qatar is guaranteed to play there.

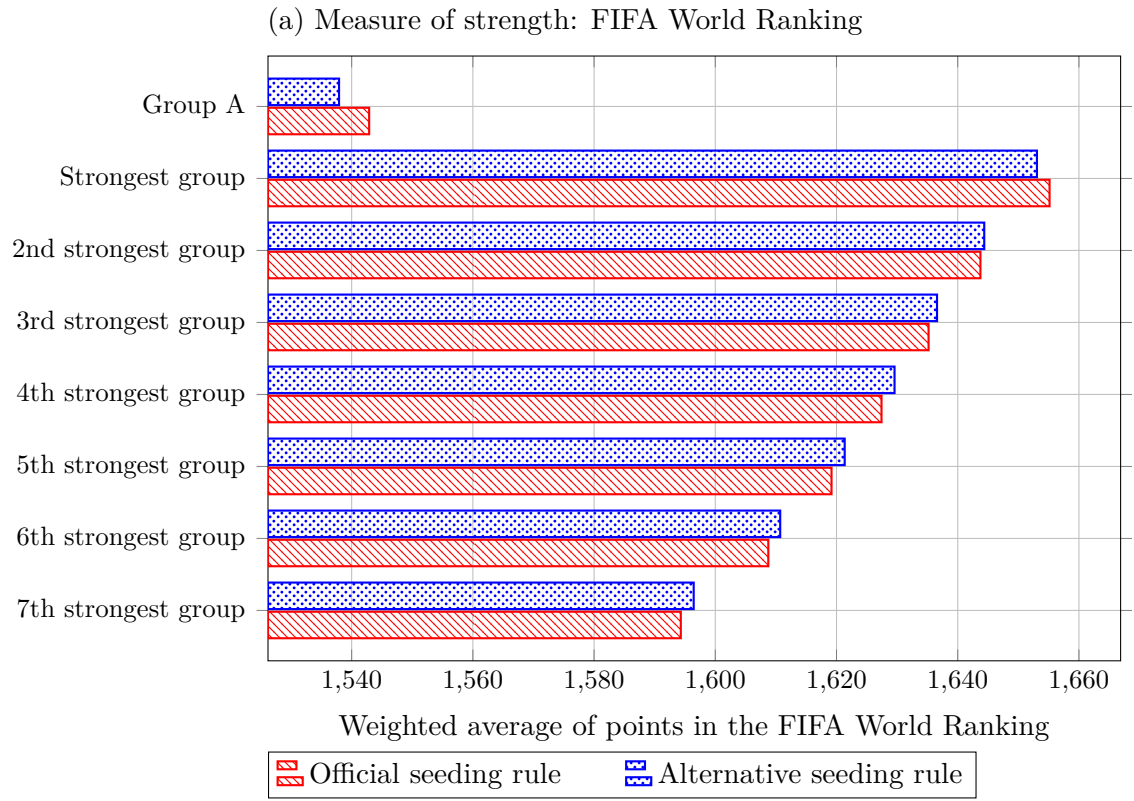
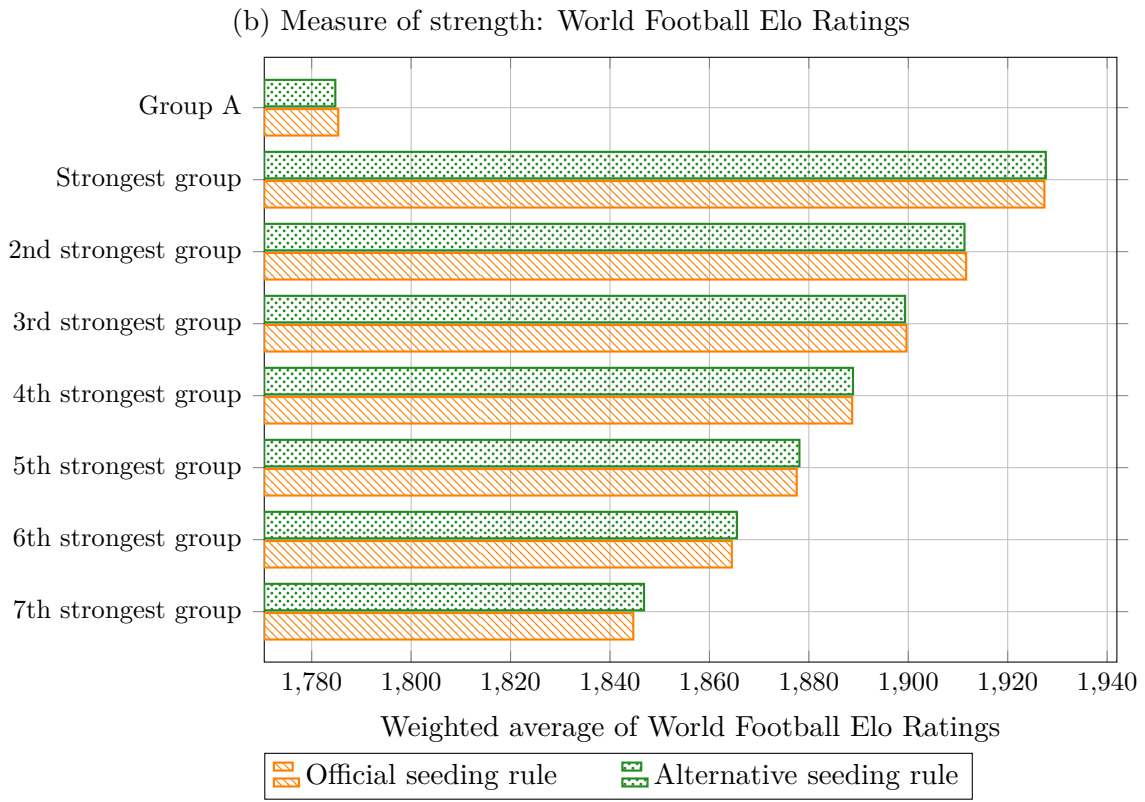
\begin{figure}[t!]
\centering

\begin{subfigure}{\textwidth}
\centering
\caption{Measure of strength: FIFA World Ranking}
\label{Fig1a}

\begin{tikzpicture}
\begin{axis}[
width = 0.8\textwidth, 
height = 0.6\textwidth,
xmajorgrids,
ymajorgrids,
scaled x ticks = false,
xlabel = {Weighted average of points in the FIFA World Ranking},
xlabel style = {align=center, font=\small},
xticklabel style = {/pgf/number format/fixed,/pgf/number format/precision=5},
xbar = 1pt,
symbolic y coords = {Group A,Strongest group,2nd strongest group,3rd strongest group,4th strongest group,5th strongest group,6th strongest group,7th strongest group},
ytick = data,
y dir = reverse,
enlarge y limits = 0.1,
legend style = {font=\small,at={(0,-0.15)},anchor=north west,legend columns=2},
legend entries = {Official seeding rule$\qquad$, Alternative seeding rule},
]
\addplot [red, thick, pattern = north west lines, pattern color = red] coordinates{
(1542.88592389571,Group A)
(1655.15444770857,Strongest group)
(1643.74817620285,2nd strongest group)
(1635.19066524714,3rd strongest group)
(1627.42479247285,4th strongest group)
(1619.17748332285,5th strongest group)
(1608.74765525428,6th strongest group)
(1594.30906858,7th strongest group)
};
\addplot [blue, thick, pattern = crosshatch dots, pattern color = blue] coordinates{
(1537.920851,Group A)
(1653.073033,Strongest group)
(1644.377188,2nd strongest group)
(1636.586913,3rd strongest group)
(1629.575912,4th strongest group)
(1621.350193,5th strongest group)
(1610.71138,6th strongest group)
(1596.434382,7th strongest group)
};
\end{axis}
\end{tikzpicture}
\end{subfigure}

\vspace{0.5cm}
\begin{subfigure}{\textwidth}
\centering
\caption{Measure of strength: World Football Elo Ratings}
\label{Fig1b}

\begin{tikzpicture}
\begin{axis}[
width = 0.8\textwidth, 
height = 0.6\textwidth,
xmajorgrids,
ymajorgrids,
scaled x ticks = false,
xlabel = {Weighted average of World Football Elo Ratings},
xlabel style = {align=center, font=\small},
xticklabel style = {/pgf/number format/fixed,/pgf/number format/precision=5},
extra x ticks = 0,
extra x tick labels = ,
xbar = 1pt,
extra x tick style = {grid = major, major grid style = {black,very thick}},
symbolic y coords = {Group A,Strongest group,2nd strongest group,3rd strongest group,4th strongest group,5th strongest group,6th strongest group,7th strongest group},
ytick = data,
y dir = reverse,
enlarge y limits = 0.1,
legend style = {font=\small,at={(0,-0.15)},anchor=north west,legend columns=2},
legend entries = {Official seeding rule$\qquad$, Alternative seeding rule},
]
\addplot [orange, thick, pattern = north west lines, pattern color = orange] coordinates{
(1785.34739014285,Group A)
(1927.37777271428,Strongest group)
(1911.64178157142,2nd strongest group)
(1899.6363,3rd strongest group)
(1888.71427671428,4th strongest group)
(1877.56198871428,5th strongest group)
(1864.50986071428,6th strongest group)
(1844.68775842857,7th strongest group)
};
\addplot [ForestGreen, thick, pattern = crosshatch dots, pattern color = ForestGreen] coordinates{
(1784.751319,Group A)
(1927.665165,Strongest group)
(1911.306196,2nd strongest group)
(1899.358132,3rd strongest group)
(1888.882309,4th strongest group)
(1878.100927,5th strongest group)
(1865.535106,6th strongest group)
(1846.842742,7th strongest group)
};
\end{axis}
\end{tikzpicture}
\end{subfigure}

\captionsetup{justification=centerfirst}
\caption{Average group strengths in the 2022 FIFA World Cup}
\label{Fig1}

\end{figure}


\begin{table}[t!]
  \centering
  \caption{Average variance of group strengths}
  \label{Table4}
\centerline{
    \rowcolors{1}{gray!20}{}
    \begin{tabularx}{1.05\textwidth}{l CCCC} \toprule \hiderowcolors
    Measure of strength & \multicolumn{2}{c}{\textbf{FIFA World Ranking}} & \multicolumn{2}{c}{\textbf{World Football Elo Ratings}} \\ 
    Seeding procedure & Official & Alternative & Official & Alternative \\ \bottomrule \showrowcolors
    Groups A--H & 1188.82 & 1233.17 & 1951.44 & 1956.95 \\
    Groups B--H & 427.78 & 377.59 & 801.08 & 764.54 \\ \bottomrule
    \end{tabularx}
}
\end{table}

Figure~\ref{Fig1} focuses on the expected difficulty levels of the groups. According to the FIFA World Ranking, the expected strength of the strongest group is reduced by our proposal, and the expected strengths of all other groups (except for Group A) are increased.
The average difficulty level of the strongest group does not change under the World Football Elo Ratings, however, the weakest groups contain better teams, thus, the groups are more balanced overall.

Consequently, the alternative seeding regime implies a smaller variance in the average strength of Groups B--H under both measures, which is underlined by Table~\ref{Table4}: our recommendation is able to reduce variance by about 10\% for the FIFA World Ranking and about 5\% for the World Football Elo Ratings. The advantage of the proposed allocation rule is rather small but it improves balance at a minimal cost, if at all.

Figure~\ref{Fig1} shows that it is substantially easier to qualify from Group A. Nonetheless, this is caused by assigning the host Qatar there, a decision not debated in the current paper. Therefore, the average variance of group strengths fro Groups A--H is essentially meaningless since it is mainly determined by the outlier Group A.

\begin{figure}[t!]
\centering

\begin{subfigure}{\textwidth}
\centering
\caption{Measure of strength: FIFA World Ranking}
\label{Fig2a}

\begin{tikzpicture}
\begin{axis}[
width = 0.8\textwidth, 
height = 0.5\textwidth,
xmajorgrids,
ymajorgrids,
xlabel = {Weighted average of points in the FIFA World Ranking},
xlabel style = {align=center, font=\small},
symbolic x coords = {1460--1470,1470--1480,1480--1490,1490--1500,1500--1510,1510--1520,
1520--1530,1530--1540,1540--1550,1550--1560,1560--1570,1570--1580,
1580--1590,1590--1600,1600--1610,1610--1620,1620--1630,1630--1640,
1640--1650,1650--1660,1660--1670,1670--1680},
xtick=data,
enlarge x limits = 0.02,
x tick label style = {rotate=90},
ymin = 0,
scaled y ticks = false,
yticklabel style = {/pgf/number format/fixed,/pgf/number format/precision=5},
ylabel = {Probability},
ylabel style = {align=center, font=\small},
legend style = {font=\small,at={(0.1,-0.4)},anchor=north west,legend columns=2},
legend entries = {Official seeding rule$\qquad$, Alternative seeding rule},
]
\addplot [red, only marks, mark = star, thick] coordinates{
(1460--1470,0)
(1470--1480,0.000475875)
(1480--1490,0.001096875)
(1490--1500,0.002846375)
(1500--1510,0.0011385)
(1510--1520,0.00781575)
(1520--1530,0.018152125)
(1530--1540,0.027487625)
(1540--1550,0.025866125)
(1550--1560,0.016305875)
(1560--1570,0.013431875)
(1570--1580,0.019835125)
(1580--1590,0.029526625)
(1590--1600,0.06105925)
(1600--1610,0.09375975)
(1610--1620,0.12920475)
(1620--1630,0.156586375)
(1630--1640,0.16001075)
(1640--1650,0.124196875)
(1650--1660,0.07251225)
(1660--1670,0.026647125)
(1670--1680,0.012044125)
};
\addplot [blue, only marks, mark = square, very thick] coordinates{
(1460--1470,0.000019625)
(1470--1480,0.000115375)
(1480--1490,0.000338125)
(1490--1500,0.003666125)
(1500--1510,0.0025495)
(1510--1520,0.0091335)
(1520--1530,0.0137955)
(1530--1540,0.03259)
(1540--1550,0.0426445)
(1550--1560,0.0180295)
(1560--1570,0.0093095)
(1570--1580,0.007348375)
(1580--1590,0.014647)
(1590--1600,0.044376375)
(1600--1610,0.096078875)
(1610--1620,0.125646875)
(1620--1630,0.14522375)
(1630--1640,0.18371275)
(1640--1650,0.1472995)
(1650--1660,0.08725425)
(1660--1670,0.016221)
(1670--1680,0)
};
\end{axis}
\end{tikzpicture}
\end{subfigure}

\vspace{0.5cm}
\begin{subfigure}{\textwidth}
\centering
\caption{Measure of strength: World Football Elo Ratings}
\label{Fig2b}

\begin{tikzpicture}
\begin{axis}[
width = 0.8\textwidth, 
height = 0.5\textwidth,
xmajorgrids,
ymajorgrids,
xlabel = {Weighted average of World Football Elo Ratings},
xlabel style = {align=center, font=\small},
symbolic x coords = {1680--1690,1690--1700,1700--1710,1710--1720,1720--1730,1730--1740,
1740--1750,1750--1760,1760--1770,1770--1780,1780--1790,1790--1800,
1800--1810,1810--1820,1820--1830,1830--1840,1840--1850,1850--1860,
1860--1870,1870--1880,1880--1890,1890--1900,1900--1910,1910--1920,
1920--1930,1930--1940,1940--1950,1950--1960,1960--1970},
xtick=data,
enlarge x limits = 0.02,
x tick label style = {rotate=90},
ymin = 0,
scaled y ticks = false,
yticklabel style = {/pgf/number format/fixed,/pgf/number format/precision=5},
ylabel = {Probability},
ylabel style = {align=center, font=\small},
legend style = {font=\small,at={(0.1,-0.4)},anchor=north west,legend columns=2},
legend entries = {Official seeding rule$\qquad$, Alternative seeding rule},
]
\addplot [orange, only marks, mark = pentagon, very thick] coordinates{
(1680--1690,0.00024925)
(1690--1700,0)
(1700--1710,0.0002355)
(1710--1720,0.001129375)
(1720--1730,0.00184775)
(1730--1740,0.00236325)
(1740--1750,0.006674125)
(1750--1760,0.013714)
(1760--1770,0.006834875)
(1770--1780,0.018175625)
(1780--1790,0.023379625)
(1790--1800,0.012060875)
(1800--1810,0.01852225)
(1810--1820,0.019348875)
(1820--1830,0.018674375)
(1830--1840,0.026383)
(1840--1850,0.041809875)
(1850--1860,0.064109625)
(1860--1870,0.07922)
(1870--1880,0.11146975)
(1880--1890,0.103791)
(1890--1900,0.12189675)
(1900--1910,0.103738)
(1910--1920,0.09238525)
(1920--1930,0.06046175)
(1930--1940,0.030373)
(1940--1950,0.015123125)
(1950--1960,0.00502075)
(1960--1970,0.001008375)
};
\addplot [ForestGreen, only marks, mark = triangle, very thick] coordinates{
(1680--1690,0)
(1690--1700,0)
(1700--1710,0.000170875)
(1710--1720,0.000411875)
(1720--1730,0.002317875)
(1730--1740,0.004100375)
(1740--1750,0.0104025)
(1750--1760,0.012717125)
(1760--1770,0.01052325)
(1770--1780,0.014763875)
(1780--1790,0.0179245)
(1790--1800,0.012696125)
(1800--1810,0.013594125)
(1810--1820,0.020388875)
(1820--1830,0.01440775)
(1830--1840,0.02588075)
(1840--1850,0.041642625)
(1850--1860,0.064229375)
(1860--1870,0.082021375)
(1870--1880,0.105641375)
(1880--1890,0.115716375)
(1890--1900,0.129279875)
(1900--1910,0.10651375)
(1910--1920,0.081498)
(1920--1930,0.059767125)
(1930--1940,0.031179125)
(1940--1950,0.014645625)
(1950--1960,0.00584825)
(1960--1970,0.00171725)
};
\end{axis}
\end{tikzpicture}
\end{subfigure}

\caption{Distribution of group strengths in the 2022 FIFA World Cup}
\label{Fig2}

\end{figure}
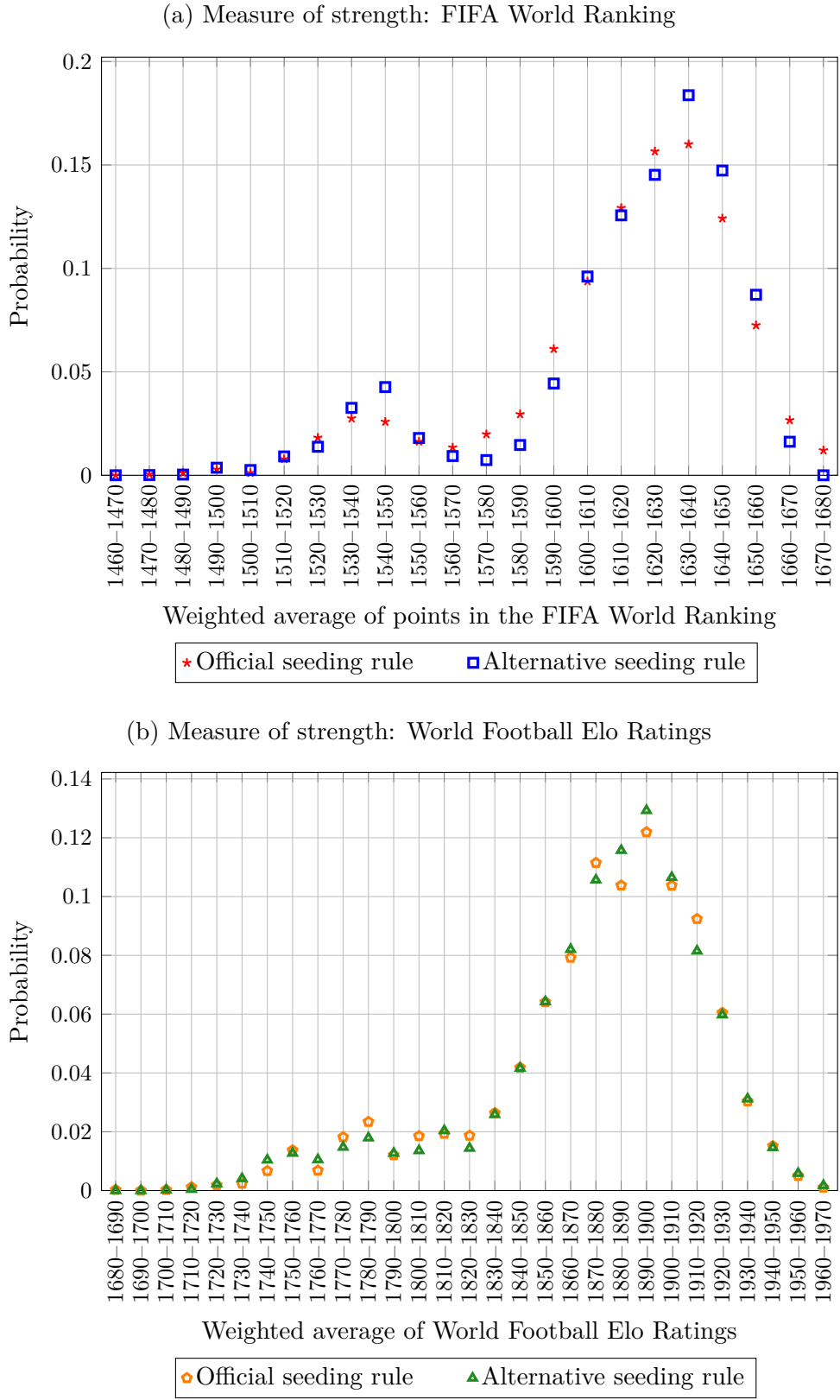


Figure~\ref{Fig2} reinforces the main message by presenting the distribution of group strengths. Clearly, the probability of a ``group of death'' is diminished if it is identified by the FIFA World Ranking. Note the case of Group A again, which accounts for having two modes. However, both distributions are more ``peaky'' around the primary mode under the alternative seeding regime, implying a more balanced level of difficulty across the groups.

\subsection{Distortions in the probability of qualification} \label{Sec42}

\begin{figure}[t!]
\centering

\begin{tikzpicture}
\begin{axis}[
width = 0.8\textwidth, 
height = \textwidth,
xmajorgrids,
ymajorgrids,
scaled x ticks = false,
xlabel = {Absolute change in percentage points},
xlabel style = {align=center, font=\small},
xticklabel style = {/pgf/number format/fixed,/pgf/number format/precision=5},
extra x ticks = 0,
extra x tick labels = ,
xbar = 1pt,
extra x tick style = {grid = major, major grid style = {black,very thick}},
symbolic y coords = {Qatar,Brazil,Belgium,France,Argentina,England,Spain,Portugal,Netherlands,Denmark,Mexico,Germany,United States,Switzerland,Uruguay,Croatia,Senegal,Iran,Japan,Morocco,Serbia,Poland,South Korea,Tunisia,Canada,Cameroon,Ecuador,Saudi Arabia,Ghana,$\mathcal{W}$/ICT PO1,$\mathcal{W}$/ICT PO2,$\mathcal{W}$/UEFA PO},
ytick = data,
y dir = reverse,
enlarge y limits = 0.02,
]
\addplot [red, thick, pattern = north west lines, pattern color = red] coordinates{
(0.0171,Qatar)
(0.0741,Brazil)
(-0.0415,Belgium)
(-0.0018,France)
(0.2885,Argentina)
(0.0311,England)
(0.0063,Spain)
(-0.0252,Portugal)
(-1.8928,Mexico)
(0.6193,Netherlands)
(0.6766,Denmark)
(0.5552,Germany)
(1.1739,Uruguay)
(0.7033,Switzerland)
(-1.7695,United States)
(0.7156,Croatia)
(0.778,Senegal)
(-0.4132,Iran)
(-0.3486,Japan)
(0.808,Morocco)
(-1.1923,Serbia)
(-1.2776,Poland)
(1.5467,South Korea)
(0.8586,Tunisia)
(-0.662,Cameroon)
(0.6084,Canada)
(-0.268,Ecuador)
(0.1851,Saudi Arabia)
(-0.3241,Ghana)
(-1.5004,$\mathcal{W}$/ICT PO1)
(0.6082,$\mathcal{W}$/ICT PO2)
(-0.537,$\mathcal{W}$/UEFA PO)
};
\end{axis}
\end{tikzpicture}

\caption{The effect of the official seeding regime compared to the alternative seeding rule on the probability of qualification in the 2022 FIFA World Cup I.}
\label{Fig3}

\end{figure}
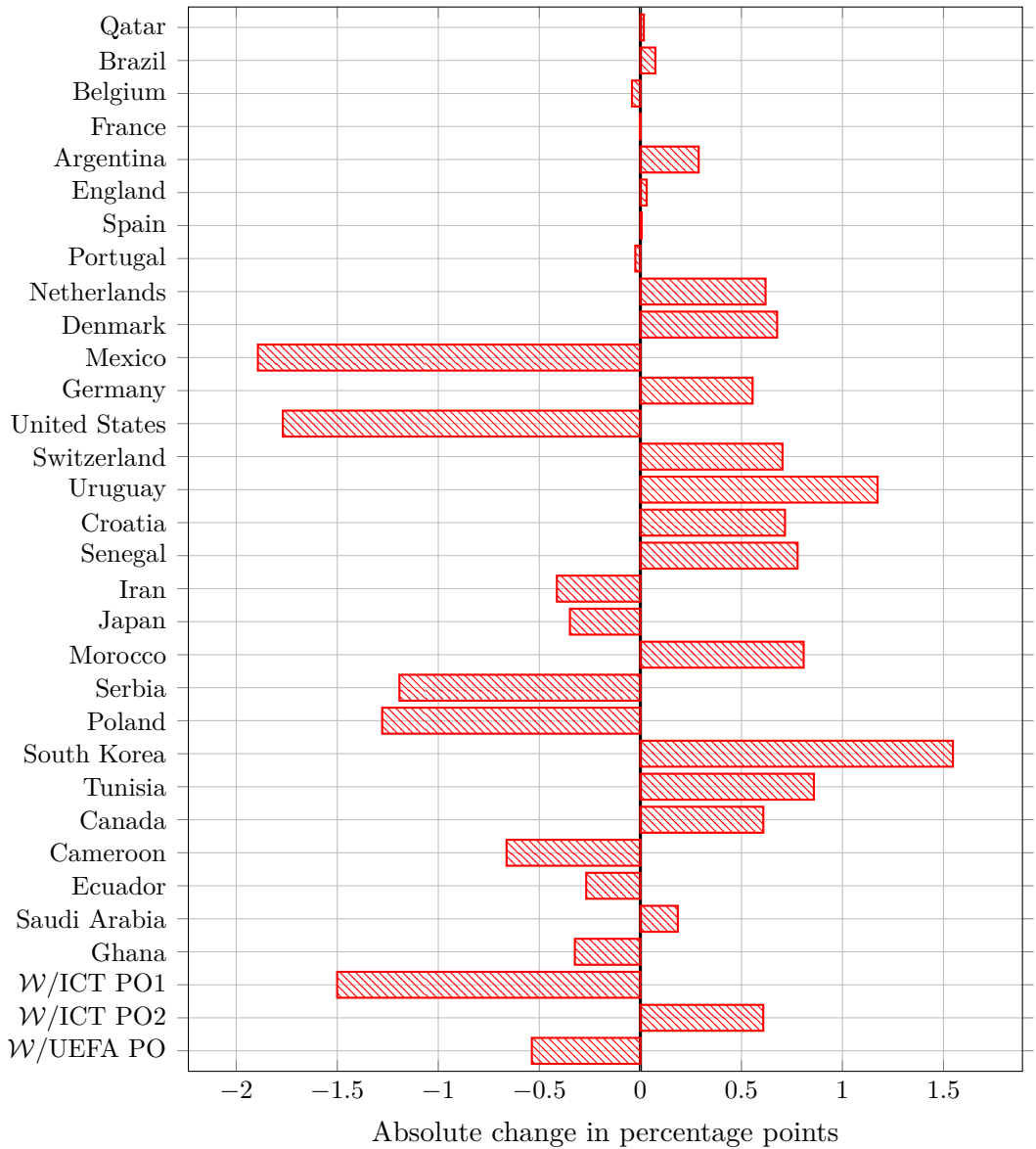


\begin{figure}[t!]
\centering

\begin{tikzpicture}
\begin{axis}[
width = 0.8\textwidth, 
height = \textwidth,
xmajorgrids,
ymajorgrids,
scaled x ticks = false,
xlabel = {Relative change in percentages},
xlabel style = {align=center, font=\small},
xticklabel style = {/pgf/number format/fixed,/pgf/number format/precision=5},
extra x ticks = 0,
extra x tick labels = ,
xbar = 1pt,
extra x tick style = {grid = major, major grid style = {black,very thick}},
symbolic y coords = {Qatar,Brazil,Belgium,France,Argentina,England,Spain,Portugal,Netherlands,Denmark,Mexico,Germany,United States,Switzerland,Uruguay,Croatia,Senegal,Iran,Japan,Morocco,Serbia,Poland,South Korea,Tunisia,Canada,Cameroon,Ecuador,Saudi Arabia,Ghana,$\mathcal{W}$/ICT PO1,$\mathcal{W}$/ICT PO2,$\mathcal{W}$/UEFA PO},
ytick = data,
y dir = reverse,
enlarge y limits = 0.02,
]
\addplot [blue, thick, pattern = crosshatch dots, pattern color = blue] coordinates{
(0.0352049608113303,Qatar)
(0.0765158344407935,Brazil)
(-0.0453033523389101,Belgium)
(-0.00189741719355263,France)
(0.333588872816493,Argentina)
(0.0350920743816596,England)
(0.00710723234218502,Spain)
(-0.03074513019099,Portugal)
(-3.55183925150167,Mexico)
(0.882655177279279,Netherlands)
(0.969671607224232,Denmark)
(0.741383684751695,Germany)
(1.72292581834941,Uruguay)
(1.04928177982537,Switzerland)
(-3.65908381256062,United States)
(1.30303124106623,Croatia)
(3.88922215556888,Senegal)
(-1.17837628189772,Iran)
(-1.13368803090812,Japan)
(3.80769267019161,Morocco)
(-2.52355715096049,Serbia)
(-3.28882504820229,Poland)
(5.23400223342696,South Korea)
(12.2890635063764,Tunisia)
(-7.8141599187894,Cameroon)
(2.05722632870988,Canada)
(-0.664794657829193,Ecuador)
(2.22698124330762,Saudi Arabia)
(-10.3971512896189,Ghana)
(-4.03326828061913,$\mathcal{W}$/ICT PO1)
(3.78846393422201,$\mathcal{W}$/ICT PO2)
(-1.24499509652769,$\mathcal{W}$/UEFA PO)
};
\end{axis}
\end{tikzpicture}

\caption{The effect of the official seeding regime compared to the alternative seeding rule on the probability of qualification in the 2022 FIFA World Cup II.}
\label{Fig4}
\end{figure}
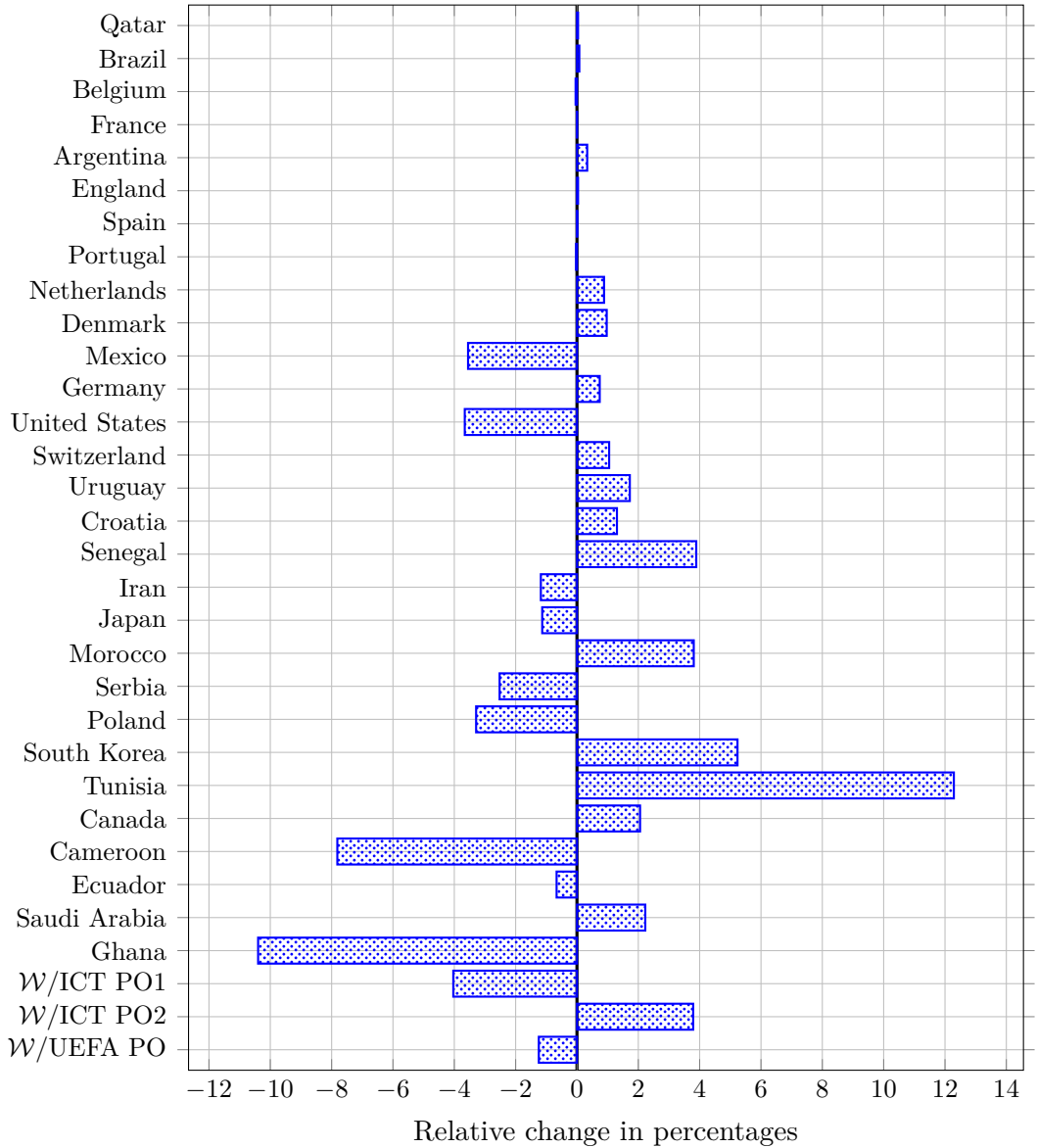


Advancement to the Round of 16 is a zero-sum game. Consequently, if there are two competitive draw procedures, one of them will favour some nations compared to the other.
Figure~\ref{Fig3} presents the effect of the official seeding rule compared to our proposal which provides a more balanced outcome as can be seen in Figures~\ref{Fig1}--\ref{Fig2}. Five countries (Mexico, the United States, Serbia, Poland, and the winner of the AFC vs CONMEBOL play-off) lose more than one percentage point in the probability of qualification. South Korea and Tunisia benefit from being assigned to Pot 3 rather than Pot 4. Uruguay is better off because the official seeding rule places two strong CONMEBOL teams (Ecuador and the winner of an inter-confederation play-off) in Pot 4 instead of only one, implying that the excepted opponent of Uruguay from Pot 4 will be weaker.

The relative effects (Figure~\ref{Fig4}) are mitigated for the teams drawn from Pots 1 and 2 but can reach or even exceed 3-4\% for weaker teams. There is a positive correlation among nations in the same association and pot: Mexico and the United States, the five European teams in Pot 2 (the Netherlands, Denmark, Germany, Switzerland, Croatia), Iran and Japan, Senegal and Morocco, Serbia and Poland, or Cameroon and Ghana. This reinforces that the results are mainly driven by the seeding policy as these teams are interchangeable in the draw.

\section{A more general comparison of seeding regimes} \label{Sec5}

So far, only the specific case of the FIFA 2022 World Cup has been examined. Therefore, it remains uncertain whether our proposal is universally advantageous concerning group balancedness. To that end, three seeding options are compared in a stylised model of the FIFA World Cup draw with an unknown placeholder, the winner of a play-off tie:
\begin{itemize}
\item
There are 33 teams;
\item
The strength of team $i$ ($0 \leq i \leq 32$) is $33-i$;
\item
One randomly chosen team is the host, automatically assigned to Pot 1 and placed in Group A;
\item
Two teams contest a play-off to be played after the draw;
\item
The teams are assigned to the pots according to their strength, except for the host and the winner of the play-off;
\item
Eight groups are formed by randomly selecting a team from each of the four pots;
\item
There are no draw constraints;
\item
Group strength is measured as before. 
\end{itemize}

We consider three options for how to seed the winner of the play-off contested by teams $i$ and $j$:
\begin{itemize}
\item
Seeding A: it is assigned according to the strength of the better team, which is equal to $\max \{ 33-i; \, 33-j \}$;
\item
Seeding B: it is assigned according to the strength of the worse team, which is equal to $\min \{ 33-i; \, 33-j \}$;
\item
Seeding C: it is assigned automatically to Pot 4.
\end{itemize}
Seeding A corresponds to our proposal for the 2022 FIFA World Cup draw. Seeding B or Seeding C can be the underlying principle of the official FIFA rule.

The seeding regimes are investigated in three different scenarios:
\begin{itemize}
\item
Setting 1: teams $i$ and $j$ are chosen randomly from the whole set ($0 \leq i,j \leq 32$) to contest the play-off, and team $i$ advances with a probability of $0.5 + (i-j)/50$;
\item
Setting 2: teams $i$ and $j$ are chosen randomly from the set of 21 weakest teams ($12 \leq i,j \leq 32$) to contest the play-off, and team $i$ advances with a probability of $0.5 + (i-j)/50$;
\item
Setting 3: teams $i$ and $j$ are chosen randomly from the set of 21 weakest teams ($12 \leq i,j \leq 32$) to contest the play-off, and team $i$ advances with a probability of $0.5 + (i-j)/25$.
\end{itemize}
In Setting 1, the ``natural'' place of the play-off winner can be in any pot. On the other hand, the participants of the undecided play-off cannot be among the 12 strongest teams according to Settings 2 and 3. The winning probabilities are more unequal in Setting 3 compared to Setting 2. For each setting, the results will be based on 1 million simulations.

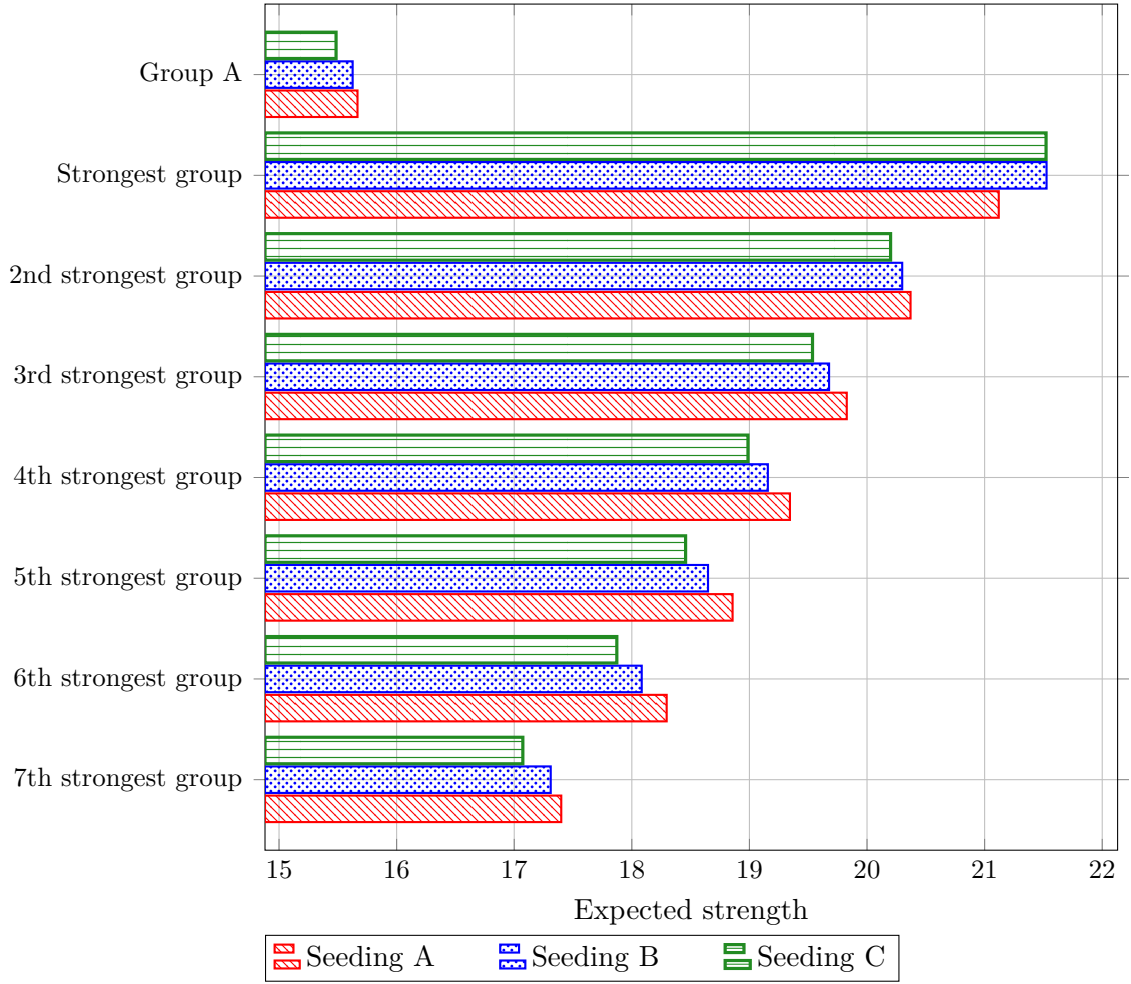
\begin{figure}[t!]
\centering

\begin{tikzpicture}
\begin{axis}[
width = 0.8\textwidth, 
height = 0.8\textwidth,
xmajorgrids,
ymajorgrids,
scaled x ticks = false,
xlabel = {Expected strength},
xlabel style = {align=center, font=\small},
xticklabel style = {/pgf/number format/fixed,/pgf/number format/precision=5},
extra x ticks = 0,
extra x tick labels = ,
xbar = 1pt,
extra x tick style = {grid = major, major grid style = {black,very thick}},
symbolic y coords = {Group A,Strongest group,2nd strongest group,3rd strongest group,4th strongest group,5th strongest group,6th strongest group,7th strongest group},
ytick = data,
y dir = reverse,
enlarge y limits = 0.1,
legend style = {font=\small,at={(0,-0.1)},anchor=north west,legend columns=3},
legend entries = {Seeding A$\qquad$, Seeding B$\qquad$, Seeding C},
]
\addplot [red, thick, pattern = north west lines, pattern color = red] coordinates{
(15.6672028571428,Group A)
(21.1201914285714,Strongest group)
(20.3700344285714,2nd strongest group)
(19.8276912857142,3rd strongest group)
(19.3445057142857,4th strongest group)
(18.8567334285714,5th strongest group)
(18.2962454285714,6th strongest group)
(17.399937,7th strongest group)
};
\addplot [blue, thick, pattern = crosshatch dots, pattern color = blue] coordinates{
(15.6259645714285,Group A)
(21.5266278571428,Strongest group)
(20.299111,2nd strongest group)
(19.6772662857142,3rd strongest group)
(19.1565888571428,4th strongest group)
(18.6478564285714,5th strongest group)
(18.0841252857142,6th strongest group)
(17.3103094285714,7th strongest group)
};
\addplot [ForestGreen, very thick, pattern = horizontal lines, pattern color = ForestGreen] coordinates{
(15.4849818571428,Group A)
(21.5224285714285,Strongest group)
(20.2003592857142,2nd strongest group)
(19.536462,3rd strongest group)
(18.9876145714285,4th strongest group)
(18.4572364285714,5th strongest group)
(17.872483,6th strongest group)
(17.0733282857142,7th strongest group)
};
\end{axis}
\end{tikzpicture}

\caption{Average group strengths, theoretical model, Setting 1}
\label{Fig5}

\end{figure}


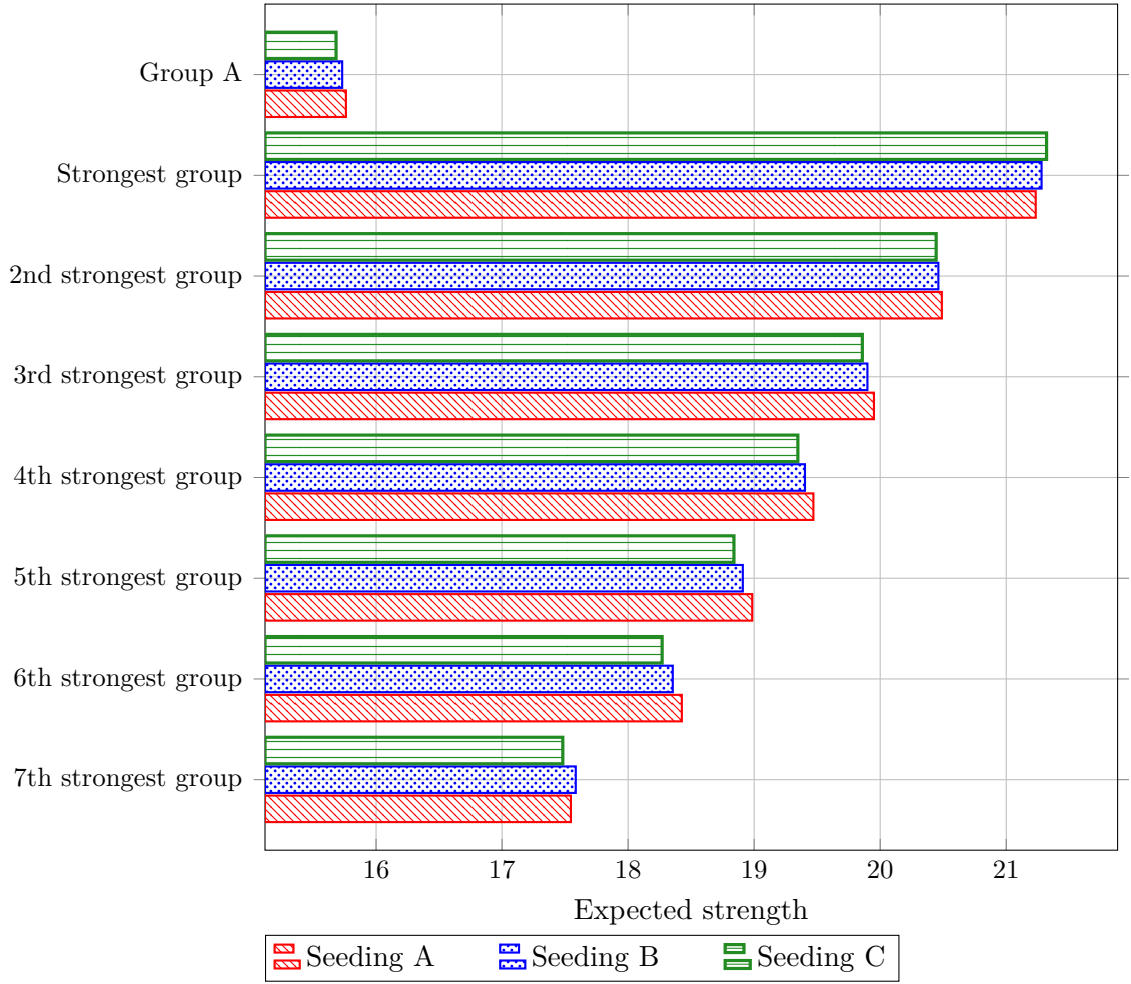
\begin{figure}[t!]
\centering

\begin{tikzpicture}
\begin{axis}[
width = 0.8\textwidth, 
height = 0.8\textwidth,
xmajorgrids,
ymajorgrids,
scaled x ticks = false,
xlabel = {Expected strength},
xlabel style = {align=center, font=\small},
xticklabel style = {/pgf/number format/fixed,/pgf/number format/precision=5},
extra x ticks = 0,
extra x tick labels = ,
xbar = 1pt,
extra x tick style = {grid = major, major grid style = {black,very thick}},
symbolic y coords = {Group A,Strongest group,2nd strongest group,3rd strongest group,4th strongest group,5th strongest group,6th strongest group,7th strongest group},
ytick = data,
y dir = reverse,
enlarge y limits = 0.1,
legend style = {font=\small,at={(0,-0.1)},anchor=north west,legend columns=3},
legend entries = {Seeding A$\qquad$, Seeding B$\qquad$, Seeding C},
]
\addplot [red, thick, pattern = north west lines, pattern color = red] coordinates{
(15.7606517142857,Group A)
(21.2338501428571,Strongest group)
(20.48903,2nd strongest group)
(19.9503085714285,3rd strongest group)
(19.4698865714285,4th strongest group)
(18.9848245714285,5th strongest group)
(18.4262935714285,6th strongest group)
(17.5461822857142,7th strongest group)
};
\addplot [blue, thick, pattern = crosshatch dots, pattern color = blue] coordinates{
(15.7312797142857,Group A)
(21.2802598571428,Strongest group)
(20.461404,2nd strongest group)
(19.8985448571428,3rd strongest group)
(19.4030944285714,4th strongest group)
(18.9099622857142,5th strongest group)
(18.354222,6th strongest group)
(17.5846455714285,7th strongest group)
};
\addplot [ForestGreen, very thick, pattern = horizontal lines, pattern color = ForestGreen] coordinates{
(15.6827364285714,Group A)
(21.3198342857142,Strongest group)
(20.4439435714285,2nd strongest group)
(19.8583212857142,3rd strongest group)
(19.3467334285714,4th strongest group)
(18.8396234285714,5th strongest group)
(18.2698672857142,6th strongest group)
(17.4818597142857,7th strongest group)
};
\end{axis}
\end{tikzpicture}

\caption{Average group strengths, theoretical model, Setting 2}
\label{Fig6}

\end{figure}


\begin{figure}[t!]
\centering

\begin{tikzpicture}
\begin{axis}[
width = 0.8\textwidth, 
height = 0.8\textwidth,
xmajorgrids,
ymajorgrids,
scaled x ticks = false,
xlabel = {Expected strength},
xlabel style = {align=center, font=\small},
xticklabel style = {/pgf/number format/fixed,/pgf/number format/precision=5},
extra x ticks = 0,
extra x tick labels = ,
xbar = 1pt,
extra x tick style = {grid = major, major grid style = {black,very thick}},
symbolic y coords = {Group A,Strongest group,2nd strongest group,3rd strongest group,4th strongest group,5th strongest group,6th strongest group,7th strongest group},
ytick = data,
y dir = reverse,
enlarge y limits = 0.1,
legend style = {font=\small,at={(0,-0.1)},anchor=north west,legend columns=3},
legend entries = {Seeding A$\qquad$, Seeding B$\qquad$, Seeding C},
]
\addplot [red, thick, pattern = north west lines, pattern color = red] coordinates{
(15.8019425714285,Group A)
(21.2436865714285,Strongest group)
(20.5026915714285,2nd strongest group)
(19.9687737142857,3rd strongest group)
(19.4929091428571,4th strongest group)
(19.0153927142857,5th strongest group)
(18.4752014285714,6th strongest group)
(17.7014482857142,7th strongest group)
};
\addplot [blue, thick, pattern = crosshatch dots, pattern color = blue] coordinates{
(15.7549957142857,Group A)
(21.3294957142857,Strongest group)
(20.4945412857142,2nd strongest group)
(19.9242297142857,3rd strongest group)
(19.4260962857142,4th strongest group)
(18.9294931428571,5th strongest group)
(18.3713908571428,6th strongest group)
(17.5978404285714,7th strongest group)
};
\addplot [ForestGreen, very thick, pattern = horizontal lines, pattern color = ForestGreen] coordinates{
(15.705571,Group A)
(21.3680527142857,Strongest group)
(20.4712381428571,2nd strongest group)
(19.8800221428571,3rd strongest group)
(19.3655745714285,4th strongest group)
(18.8546315714285,5th strongest group)
(18.2821118571428,6th strongest group)
(17.4906175714285,7th strongest group)
};
\end{axis}
\end{tikzpicture}

\caption{Average group strengths, theoretical model, Setting 3}
\label{Fig7}

\end{figure}
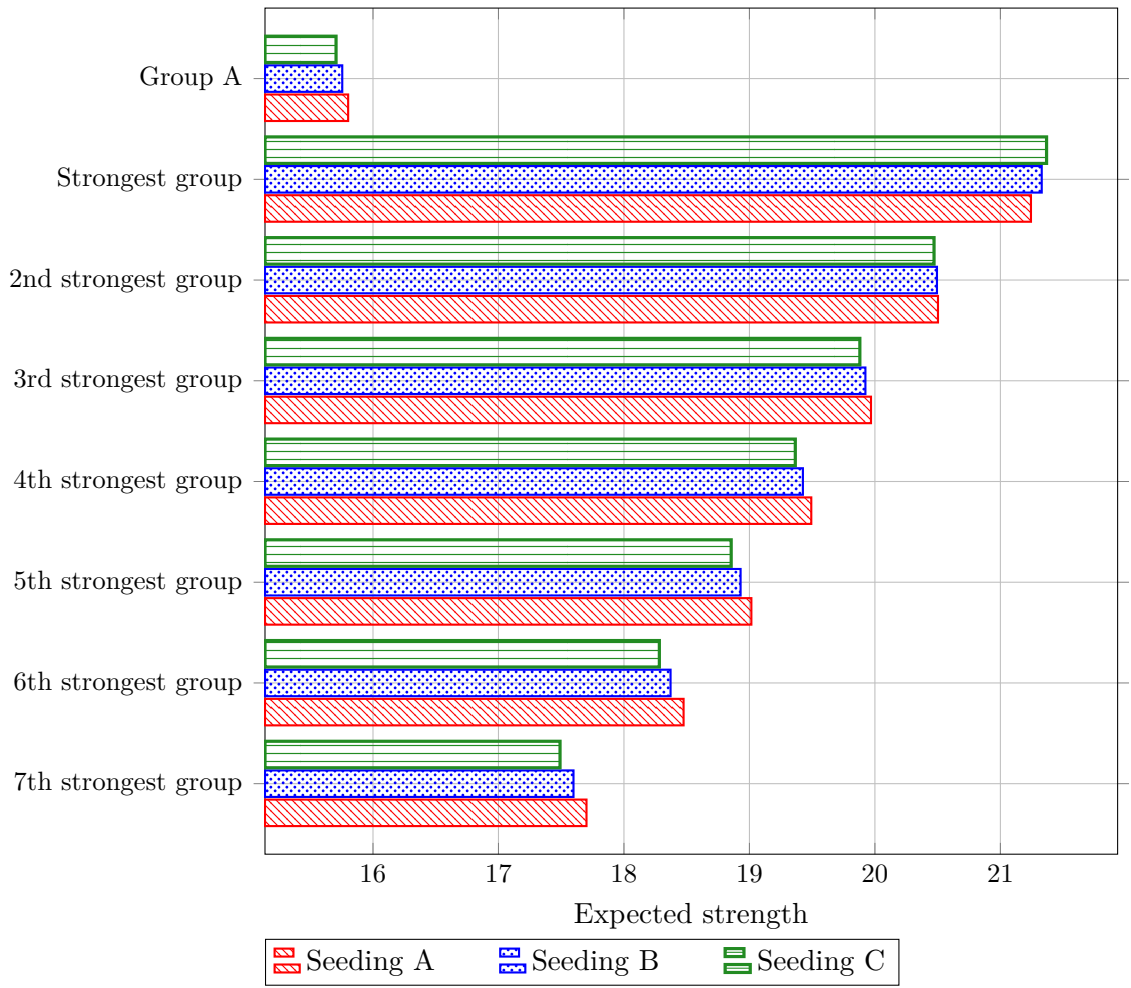


Figures~\ref{Fig5}--\ref{Fig7} show the average difficulty levels of the groups in Settings 1--3, respectively. Under Setting 1, Seeding A is the best rule to balance the groups, followed by Seeding B and Seeding C (Figure~\ref{Fig5}). In particular, the expected strength of the strongest ``group of death'' is the lowest under Seeding A, while all other groups---including Group A---are tougher according to Seeding A than according to the other two rules. This finding is intuitive: the better contestant in the play-off is the likely winner, thus, the least mistake is committed if the placeholder is assigned according to the highest-ranked possible winner. Similar to the 2022 FIFA World Cup, Group A is an outlier due to the automatic assignment of the host.

However, the situation is somewhat more complicated if the contestants of the play-off are relatively weak as in Setting 2 (Figure~\ref{Fig6}). While Seeding A minimises the imbalance across Group A and the six strongest groups, the weakest of Groups B--H is expected to be closer to the other groups under Seeding B. In other words, Seeding A allows for a relatively easy group if the play-off is won by the lower-ranked contestant which is assigned according to the strength of the higher-ranked contestant.

This conjecture is reinforced by Setting 3, where the participants of the play-off are more different in the probability of winning (Figure~\ref{Fig7}). Consequently, it is less likely, \emph{ceteris paribus}, that the lower-ranked contestant will advance from the play-off, and the pattern seen in Figure~\ref{Fig5} remains valid.

To summarise, Seeding A seems to be the best regime in general. Even though its dominance can be debated in Setting 2, the stakeholders probably prefer six balanced groups together with an easy one (after all, Group A is guaranteed to be easy by the organiser of the 2022 FIFA World Cup) rather than seven groups such that any six of them are less balanced than the six strongest groups under Seeding A.

\section{Conclusions} \label{Sec6}

The current paper has examined the draw system of the 2022 FIFA World Cup. The official seeding rule has been demonstrated to violate an important principle by failing to balance the competitive levels across the groups. Allocating the winners of the unfinished play-offs according to the highest-ranked candidate does provide a fairer outcome. The questionable decision of FIFA has harmed some countries, including Ukraine. Our proposal of using the rating of the higher-ranked team in an undecided tie for seeding purposes seems to be a fairer policy in general.

Although the methodology used to simulate the outcome of the matches played in the play-offs and the FIFA World Cup is relatively simple, we have mainly focused on the difference between the official and the alternative seeding rules. Hence, the direction of the effects (the variation in the strengths of the groups and the set of countries that benefit/suffer from the official pot allocation) are likely to remain unchanged under a wide set of prediction models.

Our study will probably inspire more researchers to analyse sports rules. Hopefully, FIFA and other tournament organisers will begin extensive consultation with the academic community before similar decisions.

\section*{Acknowledgements}
\addcontentsline{toc}{section}{Acknowledgements}
\noindent
This paper could not have been written without \emph{my father} (also called \emph{L\'aszl\'o Csat\'o}), who has primarily coded the simulations in Python. \\
We are grateful to \emph{Julien Guyon} for inspiration and useful advice. \emph{Hans van Eetvelde}, \emph{Mark Gagolewski}, \emph{Tam\'as Halm}, and \emph{D\'ora Gr\'eta Petr\'oczy} have provided important suggestions. \\
We are indebted to the \href{https://en.wikipedia.org/wiki/Wikipedia_community}{Wikipedia community} for summarising important details of the sports competition discussed in the paper. \\ 
The research was supported by the MTA Premium Postdoctoral Research Program grant PPD2019-9/2019.

\bibliographystyle{apalike}
\bibliography{All_references}

\begin{thebibliography}{}

\bibitem[Boczo\'n and Wilson, 2018]{BoczonWilson2018}
Boczo\'n, M. and Wilson, A.~J. (2018).
\newblock Goals, constraints, and public assignment: A field study of the
  {UEFA} {C}hampions {L}eague.
\newblock Technical Report 18/016, University of Pittsburgh, Kenneth P.
  Dietrich School of Arts and Sciences, Department of Economics.
\newblock
  \url{https://www.econ.pitt.edu/sites/default/files/working_papers/Working%20Paper.18.16.pdf}.

\bibitem[Cea et~al., 2020]{CeaDuranGuajardoSureSiebertZamorano2020}
Cea, S., Dur{\'a}n, G., Guajardo, M., Saur{\'e}, D., Siebert, J., and Zamorano,
  G. (2020).
\newblock An analytics approach to the {FIFA} ranking procedure and the {W}orld
  {C}up final draw.
\newblock {\em Annals of Operations Research}, 286(1-2):119--146.

\bibitem[Chater et~al., 2021]{ChaterArrondelGayantLaslier2021}
Chater, M., Arrondel, L., Gayant, J.-P., and Laslier, J.-F. (2021).
\newblock Fixing match-fixing: Optimal schedules to promote competitiveness.
\newblock {\em European Journal of Operational Research}, 294(2):673--683.

\bibitem[Csat\'o, 2022a]{Csato2022e}
Csat\'o, L. (2022a).
\newblock On the fairness of the restricted group draw problem in the 2018
  {FIFA} {W}orld {C}up.
\newblock Manuscript. {DOI}:
  \href{https://doi.org/10.48550/arXiv.2103.11353}{10.48550/arXiv.2103.11353}.

\bibitem[Csat\'o, 2022b]{Csato2022a}
Csat\'o, L. (2022b).
\newblock Quantifying incentive (in)compatibility: {A} case study from sports.
\newblock {\em European Journal of Operational Research}, 302(2):717--726.

\bibitem[Csat\'o, 2022c]{Csato2022c}
Csat\'o, L. (2022c).
\newblock Quantifying the unfairness of the 2018 {FIFA} {W}orld {C}up
  qualification.
\newblock {\em International Journal of Sports Science \& Coaching}, in press.
\newblock {DOI}:
  \href{https://doi.org/10.1177/17479541211073455}{10.1177/17479541211073455}.

\bibitem[Csat\'o, 2022d]{Csato2022b}
Csat\'o, L. (2022d).
\newblock {UEFA} against the champions? {A}n evaluation of the recent reform of
  the {C}hampions {L}eague qualification.
\newblock {\em Journal of Sports Economics}, in press.
\newblock {DOI}:
  \href{https://doi.org/10.1177/15270025221074700}{10.1177/15270025221074700}.

\bibitem[FIFA, 2018a]{FIFA2018a}
FIFA (2018a).
\newblock 2026 {FIFA} {W}orld {C}up\textsuperscript{{TM}}: {FIFA} {C}ouncil
  designates bids for final voting by the {FIFA} {C}ongress.
\newblock 10 June.
  \url{http://web.archive.org/web/20210306161039/https://www.fifa.com/who-we-are/news/2026-fifa-world-cuptm-fifa-council-designates-bids-for-final-voting-by-the-fifa-}.

\bibitem[FIFA, 2018b]{FIFA2018c}
FIFA (2018b).
\newblock Revision of the {FIFA} / {C}oca-{C}ola {W}orld {R}anking.
\newblock \url{https://img.fifa.com/image/upload/edbm045h0udbwkqew35a.pdf}.

\bibitem[FIFA, 2022a]{FIFA2022b}
FIFA (2022a).
\newblock Decisions taken concerning {FIFA} {W}orld {C}up {Q}atar
  2022\textsuperscript{{TM}} qualifiers.
\newblock 8 March.
  \url{https://www.fifa.com/tournaments/mens/worldcup/qatar2022/media-releases/decisions-taken-concerning-fifa-world-cup-qatar-2022-tm-qualifiers}.

\bibitem[FIFA, 2022b]{FIFA2022a}
FIFA (2022b).
\newblock {\em {D}raw procedures. {FIFA} {W}orld {C}up {Q}atar
  2022\textsuperscript{{TM}}}.
\newblock
  \url{https://digitalhub.fifa.com/m/2ef762dcf5f577c6/original/Portrait-Master-Template.pdf}.

\bibitem[{Football rankings}, 2020]{FootballRankings2020}
{Football rankings} (2020).
\newblock Simulation of scheduled football matches.
\newblock 28 December.
  \url{http://www.football-rankings.info/2020/12/simulation-of-scheduled-football-matches.html}.

\bibitem[Guyon, 2014]{Guyon2014c}
Guyon, J. (2014).
\newblock Rethinking the {FIFA} {W}orld {C}up\textsuperscript{{TM}} final draw.
\newblock Manuscript. {DOI}:
  \href{http://dx.doi.org/10.2139/ssrn.2424376}{10.2139/ssrn.2424376}.

\bibitem[Guyon, 2015]{Guyon2015a}
Guyon, J. (2015).
\newblock Rethinking the {FIFA} {W}orld {C}up\textsuperscript{{TM}} final draw.
\newblock {\em Journal of Quantitative Analysis in Sports}, 11(3):169--182.

\bibitem[Guyon, 2018]{Guyon2018d}
Guyon, J. (2018).
\newblock Pourquoi la {C}oupe du monde est plus \'equitable cette ann\'ee.
\newblock {\em The Conversation}.
\newblock 13 June.
  \url{https://theconversation.com/pourquoi-la-coupe-du-monde-est-plus-equitable-cette-annee-97948}.

\bibitem[Jones, 1990]{Jones1990}
Jones, M.~C. (1990).
\newblock The {W}orld {C}up draw's flaws.
\newblock {\em The Mathematical Gazette}, 74(470):335--338.

\bibitem[Kl{\"o}{\ss}ner and Becker, 2013]{KlossnerBecker2013}
Kl{\"o}{\ss}ner, S. and Becker, M. (2013).
\newblock Odd odds: The {UEFA} {C}hampions {L}eague {R}ound of 16 draw.
\newblock {\em Journal of Quantitative Analysis in Sports}, 9(3):249--270.

\bibitem[Laliena and L{\'o}pez, 2019]{LalienaLopez2019}
Laliena, P. and L{\'o}pez, F.~J. (2019).
\newblock Fair draws for group rounds in sport tournaments.
\newblock {\em International Transactions in Operational Research},
  26(2):439--457.

\bibitem[Lasek et~al., 2013]{LasekSzlavikBhulai2013}
Lasek, J., Szl{\'a}vik, Z., and Bhulai, S. (2013).
\newblock The predictive power of ranking systems in association football.
\newblock {\em International Journal of Applied Pattern Recognition},
  1(1):27--46.

\bibitem[Lasek et~al., 2016]{LasekSzlavikGagolewskiBhulai2016}
Lasek, J., Szl{\'a}vik, Z., Gagolewski, M., and Bhulai, S. (2016).
\newblock How to improve a team's position in the {FIFA} ranking? {A}
  simulation study.
\newblock {\em Journal of Applied Statistics}, 43(7):1349--1368.

\bibitem[Maher, 1982]{Maher1982}
Maher, M.~J. (1982).
\newblock Modelling association football scores.
\newblock {\em Statistica Neerlandica}, 36(3):109--118.

\bibitem[Rathgeber and Rathgeber, 2007]{RathgeberRathgeber2007}
Rathgeber, A. and Rathgeber, H. (2007).
\newblock Why {G}ermany was supposed to be drawn in the group of death and why
  it escaped.
\newblock {\em Chance}, 20(2):22--24.

\bibitem[Roberts and Rosenthal, 2022]{RobertsRosenthal2022}
Roberts, G.~O. and Rosenthal, J.~S. (2022).
\newblock Football group draw probabilities and corrections.
\newblock Manuscript. {DOI}:
  \href{https://doi.org/10.48550/arXiv.2205.06578}{10.48550/arXiv.2205.06578}.

\bibitem[Scarf et~al., 2009]{ScarfYusofBilbao2009}
Scarf, P., Yusof, M.~M., and Bilbao, M. (2009).
\newblock A numerical study of designs for sporting contests.
\newblock {\em European Journal of Operational Research}, 198(1):190--198.

\bibitem[UEFA, 2021a]{UEFA2021c}
UEFA (2021a).
\newblock {\em Regulations of the UEFA Champions League 2021-24 Cycle. 2021/22
  Season}.
\newblock
  \url{https://web.archive.org/web/20210714180923/https://documents.uefa.com/r/Regulations-of-the-UEFA-Champions-League-2021/22-Online}.

\bibitem[UEFA, 2021b]{UEFA2021e}
UEFA (2021b).
\newblock {\em Regulations of the UEFA Europa Conference League 2021-24 Cycle.
  2021/22 Season}.
\newblock
  \url{https://web.archive.org/web/20220208043024/https://documents.uefa.com/r/Regulations-of-the-UEFA-Europa-Conference-League-2021/22-Online}.

\bibitem[Van~Eetvelde and Ley, 2019]{VanEetveldeLey2019}
Van~Eetvelde, H. and Ley, C. (2019).
\newblock Ranking methods in soccer.
\newblock In Kenett, R.~S., Longford, T.~N., Piegorsch, W., and Ruggeri, F.,
  editors, {\em Wiley StatsRef: Statistics Reference Online}, pages 1--9.
  Springer, Hoboken, New Jersey, USA.

\end{thebibliography}

\end{document}